\documentstyle[a4,12pt]{article}

\begin{document}
\begin{titlepage}

\begin{flushright}
CERN-TH/97-334\\
LPTHE-ORSAY 97/66  \\
LPTM 97/51\\
hep-th/9712022 \\
\end{flushright}

\vskip.5cm
\begin{center}
{\huge{\bf Solitons in the  Horava-Witten
supergravity }}
\end{center}
\vskip1.5cm

\centerline{ E. Dudas\footnote{e.mail: edudas@cern.ch} $^{a,b}$ and
J. Mourad\footnote{e.mail: mourad@qcd.th.u-psud.fr} $^{c}$} 
\vskip 15pt
\centerline{$^{a}$ CERN-TH}
\centerline{CH-1211 Geneva 23, {\sc Switzerland}}
\vskip 5pt
\centerline{$^{b}$ Laboratoire de Physique Th\'eorique et Hautes Energies
\footnote{Laboratoire associ\'e au CNRS-URA-D0063.}}
\centerline{B\^at. 211, Univ. Paris-Sud, F-91405 Orsay Cedex, {\sc France}}
\vskip 5pt
\centerline{$^{c}$ Laboratoire de Physique Th\'eorique
et Mod\'elisation}
\centerline{Site Saint-Martin, Univ. de Cergy-Pontoise}
\centerline{F-95302 Cergy-Pontoise Cedex, {\sc France}}
\vglue .5truecm

\begin{abstract}
We study classical BPS five-brane solutions in the Horava-Witten supergravity. 
The presence
of the eleventh dimension add a new feature, namely the dependence of the
solution on this new coordinate. 
For gauge five-branes  with an instanton
size less than the eleventh radius and in the neighborhood of the center
of the neutral five-brane, important corrections to the ten-dimensional solution
appear for all values of the string coupling
constant.   
We compute the mass and magnetic charge of the five-brane solitons and
the result is shown to agree with the membrane and five-brane quantization
conditions. Compactified to four dimensions, our solutions are
interpreted as axionic strings.
   
\end{abstract}	

\vfill
\begin{flushleft}
November 1997 \\
\end{flushleft}

\end{titlepage} 

\section{Introduction}

Solitonic states  play an important r\^ole in understanding the
nonperturbative dynamics in quantum field theory
and string theory (for a review, see for ex. \cite{DKL}). In supersymmetric
theories,  BPS states are believed to be stable
under quantum corrections \cite{WO}.
Their presence is very important in checking
various duality conjectures, where solitons in one theory are mapped
into elementary particles in the dual theory.
 
It was shown by Strominger in 1990 \cite{S} that the low-energy limit of
the ten-dimensional (10d) heterotic strings have classical solutions coming
from instantons in the gauge group dressed with background values for the
gravitational multiplet. This solution turned out to be crucial for
the conjectured string-five brane duality in 10d \cite{Duff}. 

Recently, it was argued \cite{HW} that the strong coupling limit of the
$E_8 \times E_8$ heterotic string is the eleven-dimensional supergravity,
the eleventh coordinate being defined on the interval $0 \le x^{11} \le l$,
with the two gauge groups sitting on the boundaries $x^{11}=0$ and $x^{11}=l$,
respectively. The length $l$ is related to the string coupling constant
$e^{\phi_0}$ by
\begin{equation}
l={1 \over 2} (4 \pi k_{11}^2)^{1/9} e^{2 \phi_0 \over 3} \ . 
\end{equation}
  This conjecture seems to be free of gauge and gravitational 
anomalies \cite{DA} and is in agreement with constraints coming from 
membrane-five brane Dirac quantization condition \cite{DA}, \cite{BM}. 

 The purpose of this paper is the study of classical 
 five-brane solutions in the
Horava-Witten theory, thus generalizing the heterotic solutions \cite{S}
(this was also studied recently in \cite{LLO}; our results of Section 2
partly overlap with this paper). A particular
emphasis will be put on the dependence of the solutions on the eleventh 
coordinate. In particular, we will find here a new manifestation of the
phenomenon observed in \cite{W}, namely that by compactification,
if we put an instanton on one boundary, the volume of the compact space on the 
other boundary becomes smaller, with a quantity related in our case to the 
instanton size. We study this
phenomenon in more detail and show that even at weak coupling, for very
small instanton size, the Strominger solution has important corrections. 
We also consider the neutral five-brane solution and find that space-time
is effectively eleven-dimensional near the center of the five-brane even at very weak coupling.

\section{Supersymmetric five-brane solutions in Horava-Witten Lagrangian}

We look for a solution of the Horava-Witten supergravity
with six dimensional Poincar\'e  and
four dimensional Lorentz invariances and which preserves half of the supersymmetries. 
The six dimensional coordinates
parametrising the world-volume of the five-brane
are denoted by $y^a, \ a=0,\dots,5$, the four dimensional ones by $x^ \mu$
and the eleventh coordinate parametrizing $S^ 1/Z^2$ by $x^{11}$. In the following, 
we shall be interested in five-branes that are transverse to the eleventh dimension.

The most general metric with the mentioned
isometries is of the form 
\begin{equation}
g=e^{2A}dy^{a}dy^{b}\eta_{ab}+e^{2B}dx^{\mu}dx^{\nu}\delta_{\mu \nu}
+e^{2C}dx^{11}dx^{11} \ , \label{1}
\end{equation}
where $A, B$ and $C$ are functions of $x^2=x^\mu x^\nu \delta_{\mu \nu}$
and $x^{11}$. In the following all indices are raised and lowered with the 
Minkowski and Euclidean metric.
The moving basis is given by
\begin{equation}
\theta^a=e^{A}dy^a,\quad \theta^\mu=e^{B}dx^\mu,\quad
\theta^{11}=e^{C}dx^{11} \ . \label{2}
\end{equation}
The four-form field strength, in order to be invariant under
the isometries, 
has to have vanishing components where at least one of the indices is $a$.
The equation of motion for this field reads $d^*G=0$,
because $G\wedge G=0$ \footnote{The term $\int C\wedge X_8$ where $X_8=dX_7$
is a polynomial in the curvature of order $8$ modifies the equation of motion by a term in $X_8$.
This term turns out to be vanishing for the five-brane 
solution as can be checked by calculating the curvature of the the metric (1).}. 
This equation may be solved as
$G=^*d\Lambda$
where $\Lambda$ is a six-form. The most general six-form compatible with 
the isometries can be written as $\Lambda=D\epsilon_6$,
where $D$ is a function of $x^2$ and $x^{11}$ 
and $\epsilon_6$ is given by $dy^0\wedge\dots\wedge dy^{5}$.
So $G$ is determined by a single function $D$ and is given by
\begin{equation}
G=e^{-6A +2B+C}{{\epsilon^{\mu}_{\ \nu \alpha \beta}}\over{6}}\partial_\mu D
dx^{\nu} \wedge dx^{\alpha} \wedge dx^{\beta} \wedge dx^{11}+
e^{-6A-C+4B}\partial_{11} D
dx^{1}\wedge\dots\wedge
dx^{4} \ , \label{3}
\end{equation}
where $\epsilon^{\mu}_{\ \nu \alpha \beta} 
$ is the totally antisymmetric tensor
with values $\pm 1$ and $0$. In the moving basis we have
\begin{equation}
G=e^{-6A -B}{{\epsilon^{\mu}_{\ \nu \alpha \beta}}\over{6}}\partial_{\mu}D
\theta^{\nu} \wedge \theta^{\alpha} \wedge \theta^{\beta} \wedge \theta^{11}+
e^{-6A-C}{{\epsilon_{\mu\nu\alpha\beta}}
\over{24}}\partial_{11}D
\theta^{\mu}\wedge\dots\wedge
\theta^{\beta} \ . \label{4}
\end{equation}

We now demand that the solution preserve a fraction 
of the eleven-dimensional supersymmetries.
We use conventions where $\{\Gamma^I,\Gamma^J\}=2\eta^{IJ}$
and $\Gamma^{I_1 I_2\dots\ I_n}={{1}\over{n!}}(\Gamma^{I_1}\Gamma^{I_2}\dots\Gamma^{I_n}\pm$
permutations).
The supersymmetry transformation rule of the eleven dimensional gravitino is 
given by
\begin{equation}
\delta\Psi_I=D_I\epsilon+
{{\sqrt{2}}\over{288}}(
\Gamma_{I}^{\ JKLM}-8\delta_{I}^{J}\Gamma^{KLM})G_{JKLM}\epsilon \ ,
\label{5}
\end{equation}
where $G_{JKLM}$ are the components of $G$ in the moving basis.

The covariant derivative when acting on spinors reads explicitly
\begin{eqnarray}
D=d+{{1}\over{2}}[\Gamma^a\Gamma^{11}e^{-C}\partial_{11}A+\Gamma^a\Gamma^\mu
e^{-B}\partial_{\mu}A]
\theta_a\nonumber\\+{{1}\over{2}}[
\Gamma^{\mu\nu}e^{-B}\partial_{\nu}B+\Gamma^{\mu}\Gamma^{11}e^{-C}\partial_{11}B]
\theta_\mu-
{{1}\over{2}}\Gamma^{\mu}\Gamma^{11}e^{-B}
\partial_{\mu}C\theta_{11} \ . \label{7}
\end{eqnarray}

In order to impose that the solution preserve some supersymmetries we have 
to look for a solution to 
\begin{equation}
\delta\Psi_I=0.\label{5p}
\end{equation}
A straightforward calculation gives
\begin{eqnarray}
\Gamma_{a}^{\ JKLM}G_{JKLM}=
24\Gamma_a\Gamma^{5}e^{-6A-C}\partial_{11}D+24\Gamma_a\Gamma^{\mu}\Gamma^5\Gamma^{11}
e^{-6A-B}\partial_\mu D \ , \nonumber \\
\Gamma_{\mu}^{\ JKLM}G_{JKLM}=
24\Gamma^5\Gamma^{11}e^{-6A-B}
\partial_{\mu}D \ , \
\Gamma_{11}^{\ JKLM}G_{JKLM}=
24\Gamma^{11}\Gamma^{5}e^{-6A-C}\partial_{11}D \ , \nonumber \\
\Gamma^{KLM}G_{\mu KLM}=6\Gamma_\mu\Gamma^{5}e^{-6A-C}\partial_{11}D-6\Gamma^\lambda_{\ \mu}\Gamma^{5}\Gamma^{11}e^{-6A-B}\partial_\lambda D \ , \nonumber \\
\Gamma^{KLM}G_{11KLM}=-6\Gamma^\lambda\Gamma^5e^{-6A-B}\partial_\lambda D \ ,
\ \Gamma^{KLM}G_{aKLM}=0 \ . \label{10} 
\end{eqnarray}
In order to obtain these relations we have used
\begin{eqnarray}
\Gamma^{5}={{1}\over{24}}\epsilon_{\mu\nu\alpha\beta}\Gamma^{\mu\nu\alpha\beta}
\ ,\quad
\Gamma^{\mu\nu\alpha}=-\epsilon^{\mu\nu\alpha\beta}\Gamma_{\beta}\Gamma^{5}
\ ,\quad
\epsilon_{\alpha\beta\mu\nu}\Gamma^{\mu\nu}=-2\Gamma^{5}\Gamma_{\alpha\beta} \ .
\label{11}
\end{eqnarray}
We look for solutions to (\ref{5p}) of the form
$\epsilon=e^{-E}\eta$,
with $\eta$ an arbitrary constant spinor of positive
eleven dimensional chirality $\Gamma^{11}\eta=\eta$
and $E$ a function of $x^2$ and $x^{11}$.
The $a$ component of the condition (\ref{5p}) is
\begin{equation}
{{1}\over{2}}[\Gamma_a\Gamma^{11}e^{-C}
\partial_{11}A+\Gamma_a\Gamma^\mu
e^{-B}\partial_{\mu}A]
\eta+{{\sqrt{2}}\over{12}}[
\Gamma_a\Gamma^{5}e^{-6A-C}\partial_{11}D+\Gamma_a\Gamma^{\mu}\Gamma^5\Gamma^{11}
e^{-6A-B}\partial_\mu D]\eta=0, \label{12}
\end{equation}
which gives
\begin{eqnarray}
\Big(
{1\over 2}
\Gamma_a\Gamma^{11}e^{-C}
\partial_{11}A+{{\sqrt{2}}\over{12}}
\Gamma_a\Gamma^{5}e^{-6A-C}\partial_{11}D
\Big)
\eta=0 \ , \\
\Big( {{1}\over{2}}
\Gamma_a \Gamma^\mu
e^{-B}\partial_{\mu}A+
{{\sqrt{2}}\over{12}}
\Gamma_a\Gamma^{\mu}\Gamma^5\Gamma^{11}
e^{-6A-B}\partial_\mu D\Big)\eta=0 \ . \label{13}
\end{eqnarray}
The first equation implies that $\eta$ is chiral in the four dimensional space,
say $\Gamma^5\eta=\alpha\eta$, where $\alpha=\pm 1$,
and
\begin{equation}
e^{6A}\partial_{11}A+\alpha
{{\sqrt{2}}\over{6}}
\partial_{11}D=0 \ . \label{14}
\end{equation}
The solution reads
\begin{equation}
e^{6A}=-\alpha\sqrt{2}D+a \ , \label{15}
\end{equation}
where $a$ does not depend on $x^{11}$.
The second equation  gives
\begin{equation}
e^{6A}\partial_{\mu}A+\alpha{{\sqrt{2}}\over{6}}\partial_\mu D=0 \ ,
\label{16}
\end{equation}
which implies that $a$ is a constant.

The $\mu$ component of the equation (\ref{5p}) gives
\begin{eqnarray}
-e^{-B}\partial_\mu E\eta+
 {{1}\over{2}}[
\Gamma^{\mu\nu}e^{-B}\partial_{\nu}B+\Gamma^{\mu}\Gamma^{11}e^{-C}\partial_{11}B]\eta
\nonumber\\
+{{\sqrt{2}}\over{12}}[\Gamma^5\Gamma^{11}e^{-6A-B}
\partial_{\mu}D-2\Gamma_\mu\Gamma^{5}e^{-6A-C}\partial_{11}D+
2\Gamma^\lambda_{\ \mu}\Gamma^{5}\Gamma^{11}e^{-6A-B}\partial_\lambda D]\eta=0 \ . \label{17}
\end{eqnarray}
Setting all the terms to zero gives
\begin{eqnarray}
-\partial_\mu E+
{{\sqrt{2}}\over{12}}\alpha e^{-6A}
\partial_{\mu}D=0 \ , \ \partial_{\nu}B-
{{\sqrt{2}}\over{3}}\alpha e^{-6A}\partial_\nu D=0 \ , \nonumber \\
\partial_{11}B-
{{\sqrt{2}}\over{3}}\alpha
e^{-6A}\partial_{11}D=0 \ . \label{18}
\end{eqnarray}
The last two equations combined with (\ref{15}) give
\begin{equation}
\partial_{I}B={{\sqrt{2}}\over{3}}\alpha
{{\partial_{I}D}
\over{-\alpha\sqrt{2}D+a}} \ . \label{19}
\end{equation}
The solution of (\ref{19}) is
\begin{equation}
e^{3B}={{b}\over{-\alpha\sqrt{2}D+a}} \ , \label{20}
\end{equation}
where $b$ is a constant which can be set to one by a rescaling of the $x^{\mu}$,
so we get $e^{3B}=e^{-6A}$.

The $11$ component of the equation (\ref{5}) reads
\begin{eqnarray}
-e^{-C}\partial_{11}E\eta-
{{1}\over{2}}\Gamma^{\mu}\Gamma^{11}e^{-B}
\partial_{\mu}C\eta\nonumber\\
+{{\sqrt{2}}\over{12}}[\Gamma^{11}\Gamma^{5}e^{-6A-C}\partial_{11}D+2
\Gamma^\lambda\Gamma^5e^{-6A-B}\partial_\lambda D]\eta=0 \ . \label{21}
\end{eqnarray}
Setting all the terms to zero, we get
\begin{eqnarray}
-\partial_{11}E+{{\sqrt{2}}\over{12}}\alpha e^{-6A}\partial_{11}D=0 \ , \
\partial_{\mu}C-{{\sqrt{2}}\over{3}}\alpha
e^{-6A}\partial_\mu D=0 \ . \label{22}
\end{eqnarray}
The last equation combined with (\ref{15}) gives
\begin{equation}
e^{3C}={{c(x^{11})}\over{-\alpha\sqrt{2}D+a}} \ , \label{23}
\end{equation}
where $c(x^{11})$ depends only on $x^{11}$.
By reparametrisation of $x^{11}$, $c$ can be set to one so we get
$C=B$. Finally, from equations (\ref{18}) and (\ref{22}) we get  
$E=-A/2$.

In summary, for the solution to be supersymmetric
the metric has to be of the form
\begin{equation}
g=e^{2A}dy^a dy^b\eta_{ab}+e^{-4A}
(dx^\mu dx^\nu\delta_{\mu\nu}+dx^{11}dx^{11}) \ , \label{24}
\end{equation}
and the four-form $G$ be given by
\begin{equation}
G={{1}\over{(-\alpha\sqrt{2}D+a)^2}}\Big[
{{\epsilon^{\mu}_{\ \nu \alpha \beta}}\over{6}}\partial_\mu D
dx^{\nu}dx^{\alpha}dx^{\beta}dx^{11}+
\partial_{11} D
dx^{1}\wedge\dots\wedge
dx^{4}\Big] \ . \label{25}
\end{equation}
The two function $A$ and $D$ are related by (\ref{14}).

It is convenient to define $\tilde D$ by
\begin{equation}
\tilde D=
{{1}\over{-\alpha\sqrt{2}D+a}} \ , \label{26}
\end{equation}
so the expression of $G$ simplifies to
\begin{equation}
\alpha {\sqrt 2} G={{\epsilon^{\mu}_{\ \nu \alpha \beta}}\over{6}}\partial_\mu \tilde D
dx^{\nu} \wedge dx^{\alpha} \wedge dx^{\beta} \wedge dx^{11}+
\partial_{11} \tilde D
dx^{1}\wedge\dots\wedge
dx^{4} \ . \label{27}
\end{equation}
$A$ is given by $e^{6A}={\tilde D}^{-1}$,
so finally the metric is given by
\begin{equation}
g={\tilde D}^{-{{1}\over{3}}}dy^a dy^b\eta_{ab}+{\tilde D}^{{{2}\over{3}}}
(dx^\mu dx^\nu\delta_{\mu\nu}+dx^{11}dx^{11}) \ . \label{28}
\end{equation}
The Bianchi identity is also simply expressed with the aid of $\tilde D$ since
\begin{equation}
\alpha {\sqrt 2} dG=(\partial^{\mu}\partial_\mu\tilde D+\partial^2_{11}\tilde D)
dx^1\wedge\dots dx^4\wedge dx^{11}, \label{29p}
\end{equation}
so  the Bianchi identity in the bulk reads
\begin{equation}
\Delta_5\tilde D=0,\label{29}
\end{equation}
where $\Delta_5$ is the five-dimensional Euclidean Laplacian.

The gaugino transformation rule under supersymmetry is given by
$\delta\chi=\Gamma^{IJ}F_{IJ}\epsilon$.
Since the solution we are seeking is invariant under the the six-dimensional 
Poincar\'e group 
we have $F_{aI}=0$, so the transformation rule can be written as
$\delta\chi=\Gamma^{\mu\nu}F_{\mu\nu}\epsilon$.

The spinorial parameter is chiral in four dimensions, so
\begin{equation}
\delta\chi=\Gamma^{\mu\nu}F_{\mu\nu}
{{1+\alpha\Gamma^5}\over{2}}
\epsilon={{1}\over{2}}F_{\mu\nu}(\Gamma^{\mu\nu}-{{\alpha}\over{2}}
\epsilon^{\mu\nu}_{\ \ \alpha\beta}\Gamma^{\alpha\beta})\epsilon \ . 
\label{30}
\end{equation}
The vanishing of $\delta\chi$ implies that the field strength must be self or 
anti-self-dual depending on the four dimensional chirality
\begin{equation}
F_{\alpha\beta}=
{{\alpha}\over{2}}
\epsilon_{\alpha\beta}^{\ \ \mu\nu}F_{\mu\nu} \ . \label{31}
\end{equation}

In the following sections we shall examine various solutions to
the self-duality equations and the Bianchi identity subject to the Horava-Witten 
boundary conditions
for $G$. We end this section with some general remarks.

There is a class of solutions we can get by using a spinor $\epsilon$
which vary along $x^{11}$ such that it is chiral, but of opposite chirality,
on the two boundaries and non-chiral for $x^{11} \not= 0,l$. The spinor
must satisfy the Horava-Witten projection condition $\epsilon (-x^{11})
=\Gamma_{11} \epsilon (x^{11})$. Then, in the
Weyl representation, the spinor we are searching corresponds to the
following ansatz
\begin{equation}
\epsilon= e^{-E}
\left( 
\begin{array}{c} 
\cos {\pi x^{11} \over 2l} \epsilon_1 \\ 
\sin{\pi x^{11} \over 2l} \epsilon_2
\end{array} 
\right) \ , \label{310}
\end{equation} 
where $\epsilon_i$ are Majorana-Weyl spinors.
 Inserting this ansatz into (\ref{5}),
as before, it turns out that the former solution (\ref{27}), (\ref{28})
still satisfies the supersymmetry conditions $\delta \Psi_a =
\delta \Psi_{\mu}=0$. However, the supersymmetry is necessarily broken
along the eleventh dimension 
\begin{equation}
\delta \Psi_{11}= e^{-E}
\left( 
\begin{array}{c} 
-{1 \over 2l} \sin {\pi x^{11} \over 2l} \epsilon_1 \\ 
{1 \over 2l} \cos{\pi x^{11} \over 2l} \epsilon_2
\end{array} 
\right) \ 
\end{equation}
and for every value of $x^{11}$. Notice also that the spinor (\ref{310})
has a discontinuity $\epsilon (-l)=-\epsilon (l)$, which is also present in
supersymmetry breaking driven by a gaugino condensation \cite{H} or in
the Scherk-Schwarz mechanism \cite{DG}, \cite{AQ}.

 It is of some interest to rewrite the solution above (\ref{27}), 
(\ref{28}), after the Wick rotation, from
the point of view of the $4d-5d$ theory obtained after the compactification
of the six coordinates $y_a$, in the simplest
Calabi-Yau like truncation \cite{Witten}, where the truncated fields are
invariant under the $SU(3)$ holonomy group of the compactified space.
The Einstein 5d metric can be obtained through the Weyl rescaling
\begin{equation}
g^{(11)}_{ab}={\tilde D}^{-1/3} \delta_{ab} \ , g^{(11)}_{\mu \nu}=
{\tilde D}^{2/3} g^{(5)}_{\mu \nu} \ , \label{311}
\end{equation}
where here $\mu,\nu=1 \cdots 5$. 
The solution (\ref{28}) becomes in the $5d$ units simply
\begin{equation}
g={\tilde D}^{-{{1}\over{3}}}dy^a dy^b\eta_{ab}+
dx^\mu dx^\nu\delta_{\mu\nu}+dx^{5}dx^{5} \ , \label{321}
\end{equation}
hence the gravitational part of the solution becomes trivial.
The $5d$ theory contains a (universal)
hypermultiplet, which in $4d$ gives, after the Horava-Witten projection,
the dilaton $S$ containing as the real part the volume $V_6$ and as 
imaginary part the Hodge dual of the
antisymmetric tensor field $C_{5\mu\nu}$. A second complex modulus $T$ 
appears by the $5d \rightarrow
4d$ compactification, the real part of which is the fifth radius $R_5$ and
the imaginary part the pseudoscalar defined by $C_{5i \bar j}
=\epsilon_{i \bar j} b$ \cite{N},
\cite{DG}, where $i, \bar j=1,2,3$ are complex indices in compactified 
space. Then we can reexpress the solution (\ref{27}), (\ref{28}) as the
background functions
\begin{equation}
S= {\tilde D}^{-1} + i {\sqrt 2} (c-{\tilde D}^{-1}) \ , \ T=1 \ , 
\label{312}
\end{equation}
where $c$ is an arbitrary real constant corresponding to the Peccei-Quinn
symmetry $S \rightarrow S + i \alpha$. So our solution corresponds by 
compactification to an axionic string, generalizing the ones discussed in
the weakly-coupled case in \cite{R}.

\subsection{Heterotic gauge five-brane}

The simplest solution to the self-duality equations
is obtained by first chosing a $SU(2)$ subgroup of $E_8$ and putting the 
t'Hooft solution in this $SU(2)$.
In order to write the solution define $\sigma^{\mu}$ by $(1,i\alpha\vec\sigma)$, 
where $\vec \sigma$ are the Pauli matrices. Then the
solution may be written as
\begin{equation}
F=
\left({{\rho}\over{\rho^2+x^2}}\right)^2\sigma_\mu
dx^\mu\sigma^{\dagger}_\nu dx^\nu \ , \label{32}
\end{equation}
where $\rho$ is the size of the instanton and $\int tr F \wedge F = 16
\alpha \pi^2$.  Our definition is such that $tr F \wedge F = (1/30) Tr F 
\wedge F$, where $Tr$ is the trace in the adjoint of $E_8$. 

Supersymmetry invariance as well as anomaly
cancellation require that the restriction of $G$ to the boundaries be 
given by
\cite{HW}
\begin{eqnarray}
G|_{x^{11}=0}= -{1 \over \sqrt2} {k_{11}^{2/3} \over  (4 \pi)^{5/3}}
[tr(F\wedge F)-{{1}\over{2}}tr(R\wedge R)]\equiv {1 \over \sqrt 2 \alpha}
f(x^2)\epsilon_4 \ , \nonumber \\
G|_{x^{11}=l}= {1 \over \sqrt2} {k_{11}^{2/3} \over  (4 \pi)^{5/3}}
[tr(F\wedge F)-{{1}\over{2}}tr(R\wedge R)]\equiv -{1 \over \sqrt 2 \alpha}
g(x^2)\epsilon_4 \ , \label{33}
\end{eqnarray}
where $\epsilon_4 = dx^1 \wedge \cdots \wedge dx^4$ and for the one 
instanton solution given above we have
$tr(F\wedge F)=96\alpha\left({\rho}^4 /{(\rho^2+x^2)}^4 \right) \epsilon_4$.
Finally, the soliton is determined by $\tilde D$ which verifies the Laplace 
equation (\ref{29}) and is subject to the boundary conditions
\begin{equation}
\partial_{11}\tilde D|_{x^{11}=0}=f(x^2),\quad
\partial_{11}\tilde D|_{x^{11}=l}=- g(x^2) \ . \label{34}
\end{equation}

In order to solve the equation (\ref{29}) subject to the boundary conditions 
(\ref{34}) it is very convenient to  consider $\tilde D$ to be the restriction of
an even and  periodic function in $x^{11}$ with period $2l$.
Then $\tilde D$ with the given boundary conditions
satisfies the following equation
\begin{equation}
(\Delta_4 + \partial_{11}^2)\tilde D=2f\delta(x^{11})+2g\delta(x^{11}-l) \ ,
\label{35}
\end{equation}
where $\delta(x^{11})$ is the delta distribution {\it periodic}
with period $2l$:
\begin{equation}
\delta(x^{11})={{1}\over{2l}}+{{1}\over{l}}\sum_{n=1}^{\infty}
\cos{{n\pi x^{11}}\over{l}} \ . \label{36}
\end{equation}
In order to solve the equation (\ref{35}), introduce the Green function
${\cal G} (x^2, x^{11})$ defined by
\begin{equation}
(\Delta_4+\partial_{11}^{2}){\cal G}=\delta^{4}(x)\delta(x^{11}) \ . 
\label{37}
\end{equation}
From the Fourier transform of the above equation  we get
\begin{eqnarray}
{\cal G}(x,x^{11})=-{1 \over 2l} \sum_{n=-\infty}^{\infty} 
\int {d^5 k \over (2 \pi)^4}
{e^{ikx} \over k^2} \delta (k_5 - {n \pi \over l}) \ , \label{370}
\end{eqnarray}
where $x^5=x^{11}$. By performing a Poisson resummation on $n$, we can 
recast the result in the form
\begin{equation}
{\cal G}=-{1 \over{8 \pi^2}}
\sum_{n=-\infty}^{\infty}
{1 \over{[x^2+(x^{11}+2nl)^2]^{3/2}}} \ . \label{371}
\end{equation}
This form of $\cal G$ is particularly suitable to examine the behavior
for large $l$ since only the term $n=0$ contributes 
to the sum and the Green function is
that of $R^5$.  
The Fourier expansion in $x^{11}$ can be obtained by calculating the
integral in (\ref{370}); it is given by
\begin{equation}
{\cal G}=-{{1}\over{2l(2\pi)^2}}\Big( {{1}\over{x^2}}+
\sum_{n=1}^{\infty}{{2n\pi}\over{lx}}K_1(nx \pi/l)\cos{{n\pi x^{11}}\over{l}}
\Big) \ , \label{38}
\end{equation}
where $x=\sqrt{x^2}$ and $K_1$ is the McDonald function of order one. 
This form of $\cal G$ is convenient to study the limit where $x$ is much 
larger than
$l$; the asymptotic behavior of $\cal G$ in this case is given by
\begin{equation}
{\cal G}=-{{1}\over{2l(2\pi)^2}}
\Big( {{1}\over{x^2}}+{{\sqrt{2}\pi}\over{\sqrt{l} x^3}}e^{-x\pi/l}\cos{{\pi x^{11}}\over{l}}
\Big)+\dots \label{39}
\end{equation}
The first term represents the Green function on $R^{4}$.

Taking into account the condition for asymptotic flatness,
the solution to equation (\ref{35}) is given by
\begin{equation}
\tilde D= 1 +2\int d^{4}y {\cal G}(x-y, x^{11})f(y)+2\int d^{4}y 
{\cal G}(x-y,x^{11}-l) g(y) \ . \label{40}
\end{equation}
The Fourier expansion in $x^{11}$ of
$\tilde D$ can be obtained by using the form (\ref{38}) of $\cal G$.
The gauge five-brane is obtained by neclecting the term $tr  R^2$, which
is legitimate to the first order in $k_{11}^{2/3}$. 
In particular, the zero mode can be explicitly calculated and is given 
for $g=0$ and $f$ given by the instanton solution (\ref{32}), by
\begin{equation}
{\tilde D}_{10d} = 1  + {4 k_{11}^{2/3} \over l  
(4 \pi)^{5/3}} {{x^2+2\rho^2}\over
{(x^2+\rho^2)^2}} \ . \label{43}
\end{equation}
This is precisely the 10d Strominger solution expressed in M-theory
units, as it should be\footnote{Expressed in string units, the coefficient
multiplying the instanton factor in the right-hand side is equal to
$\alpha'$, in agreement with ref. \cite{DKL}.}. The relation between
${\tilde D}$ and the $10d$ dilaton is $e^{2 \phi}=((2l)^3/(4 \pi
k_{11}^2)^{1/3}) {\tilde D}$. 

The exact solution (\ref{40}) gives however, even in weak-coupling regime
$l \rightarrow 0$, important corrections to (\ref{43}) coming from 
instantons of small size. In order to 
study this quantitatively, notice that,
by using (\ref{371}) , for $\rho << l$, we can expand the expressions
\begin{eqnarray}
{\tilde D}(0,0)=1
-{{\zeta}\over{4\pi^2}\rho^3}
\int d^4y  {1 \over ({1+y^2})^4 y^3}\nonumber
\\ -{\zeta \over 2 \pi^2} \sum_{n=1}^{\infty}{ 1 \over{(2nl)^3}}
\int d^4y \left( {{1}\over{1+y^2}}\right)^4
{1 \over{\left[ {{\rho^2}\over{(2nl)^2}}
y^2+1\right]^{3/2}}} \ , \nonumber \\ 
{\tilde D}(0,l)=1
-{{\zeta}\over{2\pi^2}}\sum_{n=0}^{\infty}
{1 \over{(2n+1)^{3}l^3}}
\int d^4y \left( {1 \over{1+y^2}}\right)^4
{1 \over{[{{\rho^2}\over{(2n+1)^2l^2}} y^2+1]^{3/2}}} \ , \label{44}
\end{eqnarray}
where $\zeta=-192k_{11}^{2/3} /(4\pi)^{5/3}$ reads from (\ref{33}).  
We get
\begin{eqnarray}
{\tilde D} (0,0)=1
-{5 \pi {\zeta}\over 64 \rho^3}
-{\zeta \over{96 l^3}} (\sum_{n=1}^{\infty} {1 \over{n^3}}) +
{{3\zeta \rho^2}\over{ 16 l^5}}
(\sum_{n=1}^{\infty} {1 \over{n^5}}) +\cdots \ , \nonumber \\
{\tilde D}(0,l)=1
-{\zeta \over{12 l^3}} (\sum_{n=0}^{\infty} {1 \over{(2n+1)^3}}) +
{{\zeta \rho^2}\over{ 4 l^5}}
(\sum_{n=0}^{\infty} {1 \over{(2n+1)^5}}) +\cdots \ . \label{45}
\end{eqnarray}
By comparing (\ref{45}) with (\ref{43}), it is clear that we get 
important corrections to the Strominger solution for $x=0$ and if 
$\rho \le l$, 
i.e. in the center of the instanton and for instantons
of a size smaller than the eleventh dimension. The leading term of the
Strominger solution is ${1/l \rho^2}$, to be compared to (\ref{45}), which
shows in addition that the leading terms have different behaviour on the 
two boundaries. This means that even if the ten-dimensional solution 
gives the average in $x^{11}$ of the exact eleven-dimensional solution the
latter has important fluctuations around this average. Notice that the 
space-time near the center of the instanton is eleven-dimensional even
in the limit of small $l$. 

In the $l >>x>>\rho$ limit, the leading term in the solution behaves as
\begin{equation}
{\tilde D}=1+{(4\pi k_{11}^2)^{1/3} \over 4\pi^2(x^2+x_{11}^2)^
{3/2}} \ ,
\end{equation}
which is essentially the eleven-dimensional solution, too.

If we compactify the soliton solution to 4d, by considering the $y_a$
coordinates as compact coordinates and performing  a Wick rotation
we get an instantonic four-dimensional solution. We can define the volume of the
compactified space
\begin{eqnarray}
V_6 (x^2, x^{11}) = \int d^6 y \sqrt{g_6} = {\tilde D}^{-1} 
(x^2,x^{11}) 
V_0 \ , \label{41}
\end{eqnarray}
where $V_0$ is a reference volume. It is well known that in explicit
compactifications, the volume of compact 6-manifold defines the (inverse of)
gauge couplings. It is therefore of interest to see the dependence of the
volume on $x^{11}$ and in particular the values on the two boundaries.
By a straightforward computation we get
\begin{eqnarray}
{\tilde D}(x^2,l)-{\tilde D}(x^2,0) = {1 \over 4 \pi^2} \sum_{n=-\infty}^{\infty}
\int {d^4 y } (f-g)(y) { (-1)^n \over [(x-y)^2 + n^2 l^2]^{3/2}}
= \nonumber \\
{1 \over 2 \pi^{5/2}} \int {d^4 y} (f-g)(y) \int_0^{\infty} dt\
 t^2 e^{-(x-y)^2 t^2}
\theta_4 ({i l^2 t^2 \over \pi}) \ , \label{42}
\end{eqnarray}
where $\theta_4 (\tau) = \sum_{n= -\infty}^{\infty} (-1)^n e^{i \pi n^2 \tau}$
is the Jacobi function. We find therefore a dependence on $x^{11}$ of the
volume, in analogy with the dependence found in \cite{W}. If we put, for
example, an instanton in $x^{11}=0$, we find $V_6 (l) < V_6 (0)$ and the
gauge coupling becomes stronger on the boundary without the instanton.

\subsection{Neutral five-brane}

Consider the limit $\rho\rightarrow 0$ of the gauge five-brane. Then
then  gauge potential tends to zero but
\begin{equation}
tr(F \wedge F)\rightarrow 16 \alpha\pi^2\delta^{4}(x) \epsilon_4 \ 
\end{equation}
and if $g=0$, the equation for $\tilde D$ becomes
\begin{equation}
\Delta_5\tilde
D=-2 (4\pi k_{11}^2)^{1/3} \delta^{5}(x),
\end{equation}
which represents a neutral five-brane localised at the boundary 
in $x^{11}=0$ and $x^\mu=0$.

Neutral five-branes localised in the bulk are characterized by
\begin{equation}
f(x^2)=g(x^2)=0 \ . \label{46}
\end{equation}
In this case the function ${\tilde D}$ satisfies the equation
\begin{equation}
(\Delta_4 + \partial_{11}^2 ) \tilde D=q  
 \delta^{4}(x-x_0)
\left[ \delta (x^{11}-x^{11}_0)+ \delta (x^{11}+x^{11}_0) \right] \ ,
\label{47}
\end{equation}
where a source in $-x^{11}_0$ has been added because from its construction
$\tilde D$ must be even. 
On the other hand, the Bianchi identity is modified by a five-brane in a
way $dG=- {\sqrt 2}  k_{11}^2 \alpha T_6 \delta$  that fixes the charge
to be
\begin{equation}
q=-2 k_{11}^2  T_6 \ . \label{470}
\end{equation}
Note that $\alpha=1$ corresponds to a five-brane and $\alpha=-1$
to an anti-five-brane.
We will see in the next paragraph that this is consistent with the 
five-brane Dirac quantization  condition.

The solution of  equation (\ref{47}) is
\begin{equation}
{\tilde D}= 1 +q 
\left[ {\cal G}(x-x_0, x^{11}-x^{11}_0)+
{\cal G}(x-x_0,x^{11}+x^{11}_0) \right] \ . \label{471}
\end{equation}
In the decompactified limit $l \rightarrow \infty$, this becomes the
(symmetric under $x^{11} \rightarrow -x^{11}$) solution found in 
\cite{G} in the context of 11d SUGRA. For $l \rightarrow 0$, we recover
the 10d neutral solution \cite{S}.
For $x-x_0,x^{11}-x^{11}_0 <<l$,
\begin{equation}
{\tilde D}=1+{k_{11}^2 T_6  \over 4\pi^2} \left( {1\over [(x-x_0)^2+
(x^{11}-x^{11}_0)^2]^{3/2}}+{1\over [(x-x_0)^2+
(x^{11}+x^{11}_0)^2]^{3/2}} \right) \ .
\end{equation}
As for the gauge five-brane this means that near the center the solution
differs from its ten-dimensional limit and is basically eleven-dimensional.
In analogy with the computation we did for the heterotic five-brane, we
can compute the difference of the compactified space volume for the two
boundaries. The result is
\begin{eqnarray}
{\tilde D}(x^2,l)-{\tilde D}(x^2,0) = {q \over 8 \pi^2}
\sum_{a=\pm 1} \sum_{n=-\infty}^{\infty}  {1 \over [(x-x_0)^2 + 
(nl+a x^{11}_0)^2]^{3/2}}   \nonumber \\
= {q \over 4 \pi^{5/2}} \sum_{a=\pm 1} \int_0^{\infty} dt t^2 
e^{-(x-x_0)^2t^2 +i\pi a {x^{11}_0 \over l}} \theta 
\left[ 
\begin{array}{c} 
a{x_{11,0}\over l} \\
1/2
\end{array}
\right] ({i l^2 t^2 \over \pi}) \ . \label{48}
\end{eqnarray}
Notice that for $x^{11}_0=l/2$ we get ${\tilde D}(x^2,l)={\tilde D}(x^2,0)$,
so the difference in the two volumes really comes here from the asymmetric
position of the center of the soliton with respect to the two boundaries.

\section{Mass and charges of solitons}

We compute in the following the mass and the magnetic charge
of the soliton solutions, checking that they are compatible with the Dirac
quantization condition of the fundamental membrane-five-brane pair.
We can consider more general solutions of type $(n_1,n_2)$, in an obvious
notation. As an example, let us explicitly write the heterotic five-brane 
soliton
in the particular case of one instanton on each boundary, denoted by the
$(1,1)$ case. The solution (\ref{40}) reads
\begin{eqnarray}
{\tilde D}=1 + {4 k_{11}^{2/3} \over l  
(4 \pi)^{5/3}}
\left\{ {(x-x_1)^2 + 2 \rho_1^2 \over [(x-x_1)^2 + \rho_1^2]^2 } 
+  {(x-x_2)^2 + 2 \rho_2^2 \over [(x-x_2)^2 + \rho_2^2]^2 } \right\}
+ \cdots \ , \label{49}
\end{eqnarray}
where $x_1,x_2$ are the positions of the two instantons and the dots are
the contributions of the Kaluza-Klein modes in the expansion (\ref{38}), 
which will turn out to give no contribution in computing the mass and the
charges of the soliton solution. In the following, we use the orbifold
picture of $S^1/Z_2$ and the integrals over the eleventh dimension are 
half of the integrals on the corresponding circle $S^1$.
For the Strominger $(1,0)$ case, we get in a straightforward way
\begin{eqnarray}
Q={1 \over k_{11}} \int_{S^3 \times S^1/Z_2} G = {1 \over 2k_{11}} \int_{S^3 
\times S^1} G
= -{1 \over {\sqrt 2} \alpha}({4 \pi \over k_{11}})^{1/3} \ . 
\label{50}
\end{eqnarray}
We therefore find that the magnetic charge is the
same as in the Strominger solution \cite{S}, which is expected for a BPS
soliton.
For a more general solution $(n_1,n_2)$, the magnetic charge is 
simply multiplied by $n_1+n_2$.
Notice that the quantization condition for a four-cycle
discussed in \cite{Wi} is here automatically satisfied.

The ADM mass per unit volume \cite{DGHR} is given here by
\begin{eqnarray}
M = {1 \over 4 k_{11}^2} \int_{S^3 \times S^1} dx^{11} \sqrt{g_{11,11}}
d \Omega_3 x^3 n^\mu \left( {\partial h_{\mu \nu} \over \partial x^{\nu}} -
{\partial h_{\nu \nu} \over \partial x^{\mu}}-
\sum_{a=1}^5 {\partial h_{aa} \over \partial x^{\mu}} \right)  \nonumber \\
 \ , \label{51}
\end{eqnarray}
where $\mu, \nu$ are indices on $S^3 \times S^1$, $\Omega_3=2\pi^2$ and $x 
\rightarrow \infty$. The result is
\begin{eqnarray}
M={1 \over 2} ({4 \pi \over k_{11}^4})^{1/3} = T_6 \ , \label{52}
\end{eqnarray} 
where $T_6$ is the five brane tension. We must check that this result is
compatible with the membrane-five brane quantization condition
\begin{equation}
k_{11}^2 T_3 T_6 = n \pi \ , \ {T_3^2 \over T_6}=m \pi \ , \label{53}
\end{equation}
where $T_3$ is the membrane tension. It is readily seen that the above
expression (\ref{52}) is indeed in agreement with (\ref{53}). 
From equations (\ref{50}) and (\ref{53})
we get the following  relation between the charge and the mass of 
the soliton expressing its BPS saturation \cite{DGHR}
\begin{equation}
M=-{\alpha \over {\sqrt 2} k_{11}} Q.
\end{equation}

For the neutral five-brane with $x^{11}_0$,  the quantization condition
on $\int G$ over a
four-cycle [14] implies that the charge of the soliton $q$ must be
quantized:
\begin{equation}
q={{\alpha}\over{\sqrt{2}}}\int_{S^3\times S^1} G = 
\alpha\sqrt{2}k_{11}Q=
- (4\pi k_{11}^2)^{1/3} n,
\end{equation}
where $n$ is an integer. For $n=1$ we find that $q=-(4\pi
k_{11}^2)^{1/3}$, which is identical with the result we would get from eq.
(\ref{470}). 
The corresponding mass, computed with (\ref{51}) is
\begin{equation}
M=-{q \over 2 k_{11}^2} = T_6=-{{\alpha}\over{\sqrt{2}k_{11}}}Q \ . \label{54}
\end{equation}
The relation between the mass and the charge is the same as that
of the gauge five-brane.

\section{Conclusions}
 
The purpose of this letter is to find classical solutions in the M-theory
context of Horava-Witten and study the phenomena arising as a
consequence of the dependence of the solutions on the eleventh coordinate.
By interpreting the world-volume of the five-brane as the $6d$
compactified space, we find that the compactified space volume depends
on $x^{11}$. In the compactified theory, our classical solutions are
interpreted as axionic strings depending on the fifth coordinate.
It was shown that, for heterotic five-brane solutions, new features
appear for small (compared to the eleventh radius) instanton size,
changing significantly the Strominger solution.
We shown by an explicit computation that the mass of the heterotic and neutral
five-brane solutions are in agreement with the membrane-five-brane
quantization condition and derive the quantized charge for the neutral,
five-brane solution. 

We would like to make one comment concerning the $tr R^2$ part of the
boundary conditions (\ref{33}). 
It can be easily verified that the Riemannian curvature of our
solution (\ref{40}) has vanishing $tr(R^2)$. However from the derivation
 of the boundary conditions from ten-dimensional anomalies
 it seems that one has to use in $tr(R^2)$
not the spin connection but a connection which is the analog
of the generalised connection $\Omega_-$ which contains the spin connection
plus terms depending on $G_{11\mu\nu\lambda}$. Moreover, the spin connection 
is the one calculated in the string metric and not in the M-theory metric.
In any case, the gauge solution is obtained as in the ten-dimensional case 
by neglecting the terms $tr(R^2)$,
in order to obtain a solution to first order in $k_{11}^{2/3}$.
The discussion concerning $tr(R^2)$ is relevant for generalising the
symmetric five-brane \cite{S}. No obvious generalization is possible,
however, so this issue is not important for our purposes. 

The perspectives of the strongly-coupled $E_8 \times E_8$ heterotic
string for particle physics phenomenology seem
to be promising \cite{BD}, \cite{N}, \cite{DG}, \cite{AQ}. The classical
solutions we found in this
paper lead, by compactification, to classical solutions in 4d which may be
relevant for cosmological issues and for understanding nonperturbative
aspects of the compactified theory.


\end{document}